%% file: paper.tex
\documentstyle[12pt,epsfig,amss]{article}

\newlength{\dinwidth}                       
\newlength{\dinmargin}                      
\newcommand{\hdick}{\noalign{\hrule height1.2pt}}

\newcommand{\GeV}{{\rm GeV}}

\newcommand{\pb}{{\rm pb}}

\begin{document}

\thispagestyle{empty}  
\begin{titlepage}
\begin{flushleft}
  {\tt DESY 97-098}\hfill {\tt ISSN 0418-9833} \\
  {\tt May 1997} 
%  \\[1ex] {\bf 
%    Editors: H.-U.~Martyn, K.~Rabbertz  \\[.2em]
%    Referees: K.~Daum, Ch.~Pascaud \\[.2em]
%    Final draft, \today \\[.2em]
%    Deadline for comments: May 22, 1997, 14~h }
\end{flushleft}

\vspace*{3.0cm}
\begin{center}\begin{LARGE}
    {\bf Measurement of Event Shape Variables \\ 
      in Deep Inelastic {\boldmath $e\,p$} Scattering } \\
    \vspace*{1.cm}
    {\Large H1 Collaboration} \\
    \vspace*{2.cm}
\end{LARGE}

%================================abstract===================
%\vspace*{0.5cm}
{\bf Abstract}
\begin{quotation}
\noindent 
Deep inelastic $e\,p$ scattering data, taken with the H1 detector at HERA,
are used to study the event shape variables thrust, jet broadening 
and jet mass in the current hemisphere of the Breit frame over 
a large range of momentum transfers $Q$ between $7~\GeV$ and $100~\GeV$.
The data are compared with results from $e^+e^-$ experiments. 
Using second order QCD calculations and an approach to relate hadronisation
effects to power corrections an analysis of the $Q$ dependences of the
means of the event shape parameters is presented, 
from which both the power corrections and the strong coupling constant are
determined without any assumption on fragmentation models.  
The power corrections of all event shape variables investigated
follow a $1/Q$ behaviour and can be described by a common 
parameter $\bar{\alpha}_0$.

\end{quotation}

\vspace*{2.0cm}
{\it Submitted to Physics Letters B} %$\mbox{\boldmath $B$}$}  \\
\vfill
\end{center}
\end{titlepage}

\clearpage
\begin{flushleft}
\input{h1auts}   % authors

\bigskip\bigskip
{\it
  \input{h1inst}   % institutes
   } % end \it
\end{flushleft}

\newpage

\section{Introduction}

Hadronic final states in deep inelastic $e\,p \rightarrow e'\,X$ scattering 
(DIS neutral current interaction) offer an ideal environment to study
hadronisation phenomena and to measure the strong coup\-ling constant 
over a wide range of momentum transfer $Q$ in a single experiment.
Event shape variables have been investigated in $e^+e^-$
experiments and used to extract the strong coupling constant 
$\alpha_s(M_Z)$ independent of any jet algorithm,
see {\em e.g.} ref.~\cite{bethke}.
In deep inelastic scattering a similar analysis can be performed,
provided the current (quark) fragmentation can be isolated from the target
(proton remnant) fragmentation region.
A particularly suitable frame of reference in which
to study the current region with 
minimal contamination from target fragmentation effects is the Breit frame. 
Consider, for illustration, the quark parton model.
In the Breit system the purely space-like gauge boson $\gamma/Z$ 
with four-momentum $q_{\gamma/Z} = \{ 0,\, 0, 0, -Q \}$ 
collides with the incoming
quark with longitudinal momentum $p^{in}_{q\,z} = Q/2$.
The outgoing quark is back-scattered into the current hemisphere with 
longitudinal momentum $p^{out}_{q\,z} = -Q/2$,
while the proton fragments in the opposite hemisphere.
The available energy for fragmentation in the Breit 
current hemisphere is $Q/2$.
This situation can be compared with $e^+e^- \rightarrow q\, \bar{q}$ 
annihilation, where the available energy in one hemisphere,
{\em e.g.} defined by the thrust axis, is $\sqrt{s_{e e}}/2$.
DIS event properties in the Breit current hemi\-sphere should be similar and
may be comparable to those in $e^+ e^-$ collisions at a 
scale $Q = \sqrt{s_{e e}}$.

In this paper an analysis of the event shape variables thrust, jet broadening
and jet mass in the current hemisphere of the Breit frame 
in deep inelastic $e\,p$ scattering is presented and
compared with results from $e^+e^-$ experiments. 
The kinematic phase space extends over $7~\GeV < Q < 100~\GeV$ in momentum
transfer and $0.05 < y < 0.8$ in the kinematic variable $y$.
Based on only recently available second order QCD 
calculations~\cite{mepjet,disent}
and on new developments in the treatment of non-perturbative hadronisation
contributions to final states in 
deep inelastic scattering~\cite{webber},
the $Q$ dependences of the mean event shape parameters are investigated to
study simultaneously both the power, or hadronisation, corrections
and the strong coupling constant $\alpha_s(M_Z)$ 
independently of any fragmentation model.

\section{Data Selection and Analysis Procedure}
\label{analysis}

\paragraph{The H1 Detector}
Deep inelastic $e\,p$ scattering events were collected
with the H1 detector %~\cite{h1det} 
at HERA ($E_e = 27.5~\GeV$, $E_p = 820~\GeV$, $\sqrt{s} = 300~\GeV$).
The H1 detector is described elsewhere~\cite{h1det};
here only the components relevant for the present analysis are briefly
recalled.
Only calorimetric information is used to reconstruct
electromagnetic and hadronic clusters.
The event vertex and the direction of the scattered lepton are obtained with
the tracking detectors.

The calorimeter system consists of a liquid argon (LAr) calorimeter, a backward
calorimeter and a tail catcher (instrumented iron yoke).
The LAr sampling calorimeter covers the polar angle\footnote{Polar angles
  $\theta$ are defined with respect to the incident proton direction.}
range 4$^{\circ} \le \theta \le$ 153$^{\circ}$ and all azimuthal
angles.
It consists of a lead/argon electromagnetic section  
followed by a stainless steel/argon section
for the measurement of hadronic energy.
Electromagnetic energies are measured with a resolution
of $\sigma(E)/E \simeq 12\% / \sqrt{E}\oplus1\%$ and hadronic energies
with $\sigma(E)/E \simeq 50\%/ \sqrt{E}\oplus2\%$,
as obtained from test beam results.
The absolute energy scales are known to 3\% and 4\% for
electrons and hadrons, respectively.
A lead/scintillator electromagnetic backward calorimeter (BEMC) extends the
coverage at large angles
(155$^{\circ} \le \theta \le$ 176$^{\circ}$).
The electromagnetic energy calibration is known to 1\%.
Since 1995 the backward region has been equipped with a lead/scintillating
fibre calorimeter~\cite{spacal}. 
The instrumented iron flux return yoke is used to
measure the leakage of hadronic showers.

Located inside the calorimeters is a tracking system, which consists of
central drift and proportional chambers
        (25$^{\circ} \le \theta \le$ 155$^{\circ}$),
a forward track detector
        (7$^{\circ} \le \theta \le$ 25$^{\circ}$)
and a backward proportional chamber
        (155$^{\circ} \le \theta \le$ 175$^{\circ}$).
In 1995 the latter was replaced by backward drift chambers.
The tracking chambers and calorimeters are surrounded
by a superconducting solenoid providing a uniform field of
1.15~T throughout the tracking volume.

The luminosity is determined from the rate of the Bethe--Heitler process
$e\, p \rightarrow e\, p\, \gamma$ measured in a luminosity 
monitor far downstream in the electron direction.

\paragraph{Data Selection}
The DIS data cover a large range of momentum transfer 
$Q$ between $7~\GeV$ and $100~\GeV$
and are selected with the following criteria:
\begin{enumerate}
  \item The energy of the isolated scattered lepton has to exceed
    $E'_e  >  10~\GeV$.
  \item The polar angle of the scattered lepton has to be within
    $157^\circ  <  \theta_e  <  173^\circ$ 
    (low $Q$ sample) 
    or
    $30^\circ <  \theta_e  <  150^\circ$ 
    (high $Q$ sample).
  \item The hadronic energy clusters have to be well contained within the
    calorimeter acceptances
    $5.7^\circ < \theta_{h} < 170^\circ$.
  \item There must be hadronic energy in the forward region 
    $(4^\circ <\theta < 15^\circ)$
    which is larger than $0.5~\GeV$
    in order to exclude diffractive events with large rapidity gaps.
  \item The energy and longitudinal momenta must satisfy
    $30~\GeV < \sum_i\,E_i\,(1 - \cos\theta_i) < 65~\GeV$,
    where the sum extends over all energy clusters.
  \item The total hadronic energy %$\sum_h \, E^c_h$
    in the Breit current hemisphere has to exceed 
    $0.1 \, Q$.
    %$\sum_h \, E^c_h > 0.1 \, Q$.
  \item The kinematic variables $y$ have to fulfill
    $0.05  <  y_e  <  0.80$ (measured using the lepton)
    and $0.05 <  y_h$ (measured using the hadronic energy flow).
\end{enumerate}
 
For the low $Q$ event sample 
($Q = 7 - 10~\GeV$, 1994 data with $e^+$ beam, 
integrated luminosity ${\cal L} \simeq 2.9~\pb^{-1}$)
the lepton is detected in the BEMC.
The lepton direction is measured using the event vertex and 
the backward proportional chamber in front of the calorimeter. 
For the high $Q$ event sample 
($Q = 14 - 100~\GeV$, 1994 -- 1996 data with $e^\pm$ beams, 
${\cal L} \simeq 11.7~\pb^{-1}$)
the lepton is detected in the LAr calorimeter
and its direction is taken from an associated track measured in the central 
tracking system. 

The event selection criteria are chosen so as to ensure a good measurement 
of the final state and to provide a clean DIS data sample.
Requirements 1, 2 and 7 assure that the kinematic quantities are well measured.
The criterion 5 provides a good containment of the final state particles and
rejects events with a hard photon radiated from the incoming lepton and
together with the high $y$ cut of requirement 7 suppresses
photoproduction events with a misidentified lepton.
Using tagged events, the photoproduction background 
is estimated to be about 3\% in the low $Q$ sample and to be neglible at 
high values of $Q$.
To define shape variables, events are required to have hadronic activity in 
the Breit current hemisphere according to criterion 6.
%The criterion 6 requires a minimum activity in the Breit current hemisphere
%which can be empty and thus an event shape variable cannot be defined
%(see below).

\paragraph{Analysis Procedure}
The kinematic quantities needed to perform the Breit frame transformation, 
the negative squared momentum transfer $Q^2$ and Bjorken $x = Q^2/y\,s$, 
are calculated from the scattered lepton ($E'_e, \, \theta_e, \, \phi_e$)
and hadron ($E_h, \, \theta_h, \, \phi_h$) measurements using
\begin{eqnarray}
      Q^2 & = & 4\,E_e\,E'_e\,\cos^2\frac{\theta_e}{2}\ ,  \\[.2em]
      y \ \equiv \ y_e & = & 1 - \frac{E'_e}{E_e}\,\sin^2\frac{\theta_e}{2}  
      \quad\quad\quad\quad {\rm for} \quad  y_e \, > \, 0.15  \ ,    \\[.2em]
      y \ \equiv \ y_h & = & \frac{\sum_h E_h\,(1-\cos\theta_h)}{2\,E_e} 
      \quad\quad\, {\rm for} \quad  y_e \, < \, 0.15 \ .
\end{eqnarray}
The momentum transfer $Q^2$ is always best measured using the lepton.
The kinematic variable $y$ is taken from the lepton %measurement
for sufficiently large values.
However, since the resolution in $y_e$ degrades severely at low values,
$y$ is determined using $y_h$ if $y_e < 0.15$.
This procedure ensures least uncertainty in the Lorentz transformation
to the Breit frame.

The data are corrected 
for detector effects (acceptance, resolution, energy scale)
using a full Monte Carlo simulation of the detector response.
The Monte Carlo event selection and analysis are the same as for the data.
The LEPTO event generator~\cite{lepto} 
including matrix elements plus parton 
showers is used over the full $Q$ range.
At low $Q$, however, events generated with ARIADNE~\cite{cdm}
according to the colour dipole model are also included in order to enlarge the 
Monte Carlo statistics. 
Both simulations are based on the MRS~H parton density 
parametrisations~\cite{mrsh} and they describe the data well.

The observed differential event shape distributions $\{ F_i \}$
are corrected with a matrix method 
$ F_i^{cor} = \sum_j\,C_{i j}\,F_j^{data}$.
In principle the matrix ${\cal C}$ may be obtained by an inversion of the
matrix ${\cal M}$ which transforms the `true' values into observed values
$F_i^{obs} = \sum_j\,M_{i j}\,F_j^{true}$,
including the detector effects as described by a Monte Carlo simulation.
In practice, however, the inversion of ${\cal M}$ poses mathematical problems
and leads to instabilities and oscillations,
unless extremely large Monte Carlo statistics are available.
Instead, the probability densities $\rho_{i j}^{MC}$ correlating the
observed and `true' quantities, {\em i.e.} 
$ F_i^{true} = \sum_j\,\rho_{i j}^{MC}\,F_j^{obs}$,
are used to get the correction matrix 
$ C_{i j} = \rho_{i j}^{MC}\,/\,\sum_k\,\rho_{k j}^{MC}$.
This method may still have some model dependence, which could be overcome by an
iterative procedure. 
However, as the data are described acceptably by the
Monte Carlo simulations no iteration is performed.

Systematic uncertainties in the data analysis are evaluated
for the mean values of the event shape variables, which are computed from the
differential distributions.
The following sources are considered:
(i) the influence of the unfolding procedure,
{\em i.e.} applying the matrix method or a bin-by-bin correction method; 
(ii) a variation of the energy scales,
which affects the Breit frame transformation 
but has a small effect on the event shape parameters, since they
involve ratios of hadronic momenta or energies;
and (iii) the use of the HERWIG event generator~\cite{herwig}
with fragmentation properties which are different from LEPTO.
All contributions are of similar size over the whole range of $Q$.
QED radiative corrections, which are studied with the DJANGO event 
generator~\cite{django},
are found to have a negligible effect on the event shapes.
The sum of all systematic errors of the mean event shape variables varies
between
$\sim 2\%$ at low $Q$ values and up to $\sim 8\%$ at high $Q$ values.

\section{Event Shapes in the Breit Current Hemisphere}

The infrared safe event shape variables thrust $T_c$ and $T_z$, the jet
broadening $B_c$ and the jet mass $\rho_c$ 
are studied in the Breit current hemisphere.
The definitions of the event shape variables are given below,
where the sums extend over all hadrons $h$
(being a calorimetric cluster in the detector or a parton in the QCD
calculations)  
with four-momentum 
$p_h = \{ E_h,\, {\bf p}_h \}$ fulfilling
$\cos\, (\, {\bf p}_h\cdot {\bf n} \, ) > 0$.
The current hemisphere axis ${\bf n}  = \{0,\,0,\,-1 \}$
coincides with the virtual boson direction.

\begin{itemize}
  \item {\bf Thrust {\boldmath $T_c$}}
    \begin{eqnarray}
      T_c & = & \max \, \frac{\sum_h |\, {\bf p}_h\cdot {\bf n}_T \, |} 
                 {\sum_h |\, {\bf p}_h \, |} 
      \quad\quad\quad\quad\quad\quad\quad\quad \ 
      {\bf n}_T \ \equiv \ \mbox{thrust axis} \ ,
      \quad \ \
      \label{thrust}
    \end{eqnarray}
  \item {\bf Thrust {\boldmath $T_z$}}
    \begin{eqnarray}
      T_z & = & \frac{\sum_h |\, {\bf p}_h\cdot {\bf n} \, |} 
                 {\sum_h |\, {\bf p}_h \, |} 
          \ = \ \frac{\sum_h |\, {\bf p}_{z\,h}\, |}
                     {\sum_h |\, {\bf p}_h \, |}
      \quad\quad\quad\quad \ \ \
      {\bf n} \ \equiv  \ \mbox{hemisphere axis} \ ,
      \label{thrustz}
    \end{eqnarray}
  \item {\bf Jet Broadening {\boldmath $B_c$}}
    \begin{eqnarray}
        B_c & = & \frac{\sum_h |\, {\bf p}_h\times {\bf n} \, |} 
                     {2\,\sum_h |\, {\bf p}_h \, |}
          \ = \ \frac{\sum_h |\, {\bf p}_{\perp\,h}\, |}
                     {2\,\sum_h |\, {\bf p}_h \, |}
      \quad\quad\quad\quad 
      {\bf n} \ \equiv \ \mbox{hemisphere axis} \ ,
      \label{bparameter}
    \end{eqnarray}
  \item {\bf Jet Mass {\boldmath $\rho_c$}}
    \begin{eqnarray}
      \rho_c & = & \frac{M^2}{Q^2} 
      \ = \ \frac{(\, \sum_h \, p_h \, )^2}{Q^2} \ .
      \phantom{xxxxxxxxxxxxxxxxxxxxxxxxxxxxxxx}
      \label{jetmass}
    \end{eqnarray}
\end{itemize}

\begin{figure}[p] \centering \unitlength 1mm
  \begin{picture}(170,200)
    \put(-5,0)  
     {\mbox{
       \epsfig{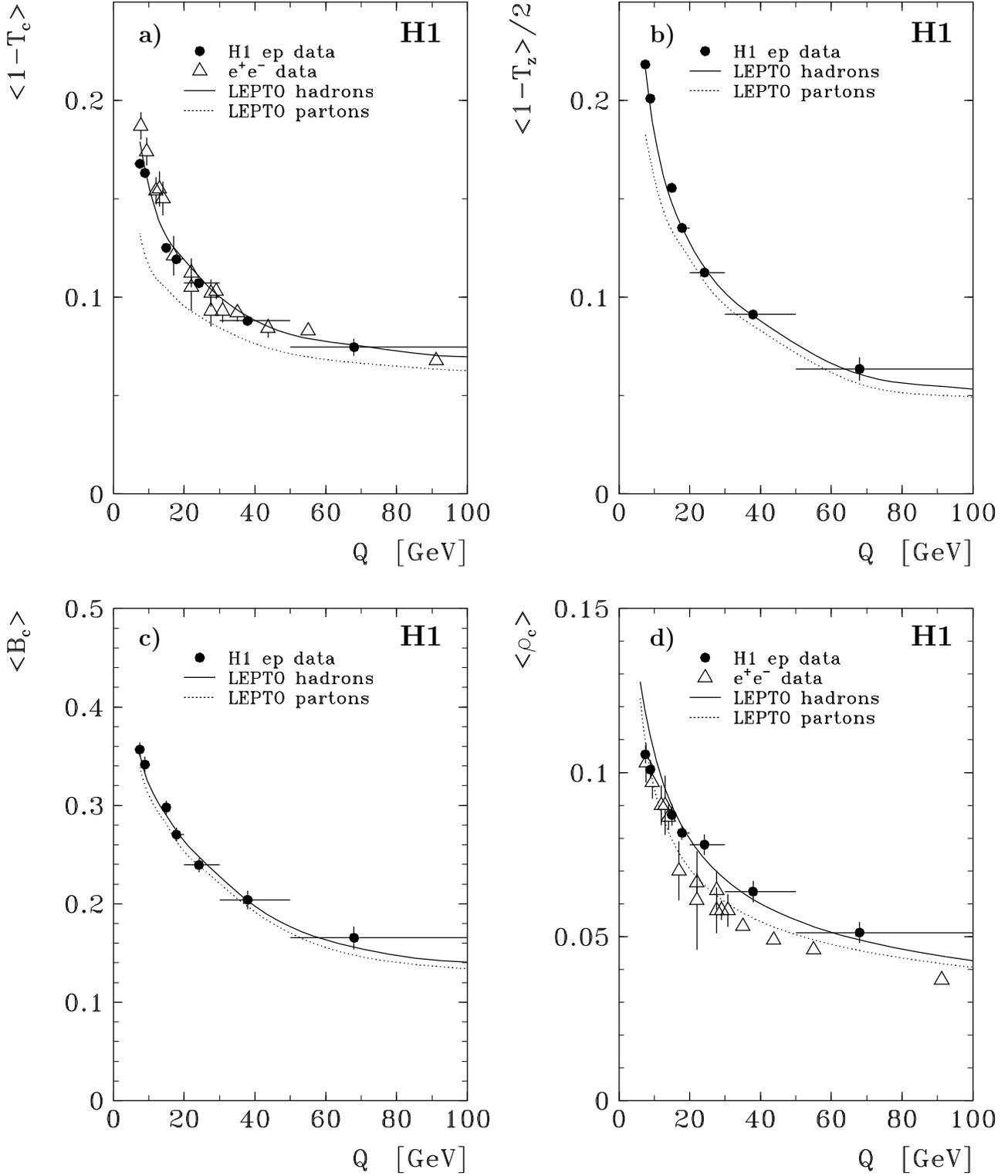} } } 
    \put( 22,197.5) {\bf a) }
    \put(110,197.5) {\bf b) }
    \put( 22, 93.5) {\bf c) }
    \put(110, 93.5) {\bf d) }
    \put( 67,197.5) {\large\bf H1 }
    \put(155,197.5) {\large\bf H1 }
    \put( 67, 93.5) {\large\bf H1 }
    \put(155, 93.5) {\large\bf H1 }
  \end{picture}
  \caption{Mean event shape variables as a function of Q for
           {\bf a)} $\langle 1 - T_c \rangle$,
           {\bf b)} $\langle 1 - T_z \rangle / 2$,
           {\bf c)} $\langle B_c \rangle$, and
           {\bf d)} $\langle \rho_c \rangle$.
            H1 DIS $e\,p$ data ($\bullet$, errors include statistics
            and systematics) 
            are compared with LEPTO Monte Carlo simulations for hadrons (---)
            and partons ($\cdot\cdot\cdot$).
            Data from $e^+e^-$ experiments ($\triangle$) are shown for
            thrust $\langle 1 - T_{ee} \rangle$, calculated for the whole
            event, and for the average of the heavy and light jet masses}
  \label{meandata}
\end{figure}

The normalized differential spectra of the event shape variables in the current
hemisphere of the Breit frame and their mean values as a function of $Q$
are presented in figs.~\ref{cdistr} and \ref{meandata}.
The mean values of the event shape parameters %as a function of $Q$ 
are also listed in table~\ref{means}. 
They extend over a large range of momentum transfers $Q$ between
$7~\GeV$ and $100~\GeV$.
The corrected data, spectra and mean values,
are well described by the LEPTO Monte Carlo generator for all $Q$.

A common characteristic of the event shape spectra of
$1 - T_c, \ 1 - T_z, \ B_c$ and $\rho_c$ is that they get narrower and more
peaked towards lower values as $Q$, or the available energy in the Breit 
current hemisphere, increases.
This means that the energy flow becomes more collimated along the event shape
axis.
This fact is also evident from the mean values of the 
event shape variables, which exhibit a clear decrease with rising $Q$.

The energy dependence of the mean thrust and jet masses 
of the H1 DIS $e\,p$ data may be compared with
available $e^+e^-$ results~\cite{eedata} at the scale $Q = \sqrt{s_{e e}}$.
Recall, however, that the event shape parameters in $e\,p$ scattering are
calculated in a single hemisphere, 
{\em i.e.} the current hemisphere of the Breit system,
and that a `natural' axis comparable with the virtual boson direction does not 
exist in $e^+e^-$ annihilation. 
There, the virtual boson is at rest and
the event axis is typically chosen to be the thrust axis.

The $Q$ dependence of the means $\langle 1 - T_c \rangle$ 
in fig.~\ref{meandata}a
is in qualitative agreement with the $e^+e^-$ measurements.
However, the $e^+e^-$ data points, coming from many experiments,
tend to lie systematically above the $e\,p$ data 
at low values of $Q \lesssim 15~\GeV$.
This can be partially understood from the fact that in $e^+e^-$
experiments thrust is calculated for the whole event and not in a 
single hemisphere, which tends to deplete the high thrust region.
This property is also demonstrated 
by comparing the thrust spectra 
of fig.~\ref{cdistr}a with the corresponding distributions at similar energies
of the PLUTO and TASSO experiments~\cite{eedata}.
Furthermore, from the definition of eq.~(\ref{thrust}), it can be seen
that thrust $T_c$ measures a mixture of the longitudinal and transverse 
momentum components with respect to the boson direction.
In strongly collimated events the thrust axis is close to the $\gamma/Z$
axis and $T_c$ is essentially given by the longitudinal momenta.
However, for more isotropic events, which frequently occur at low $Q$,
the thrust axis flips by up to $90^\circ$ and $T_c$ is dominated by
transverse momenta, resulting in higher thrust values and thus mimicking
jet-like event configurations.
Such a situation does not occur in $e^+e^-$ analyses.
The thrust definition $T_z$ of eq.~(\ref{thrustz}) is expected to be closer to
the $e^+e^-$ case. 
Indeed, with a proper rescaling in order to cover the same thrust range,
the mean values $\langle 1 - T_z \rangle /2$
exhibit a stronger $Q$ dependence, see fig.~\ref{meandata}b, 
and are comparable to $\langle 1 - T_{ee} \rangle$ at $Q \lesssim 50~\GeV$. 
%see fig.~\ref{meandata}b.

The mean jet masses $\langle \rho_c \rangle$ in fig.~\ref{meandata}d
show a weaker $Q$ dependence for $e\,p$ scattering than 
for $e^+e^-$ collisions, 
where the average of the heavy and light jet masses of both event hemispheres
are plotted.
The $e\,p$ data are systematically higher at larger values of
$Q \gtrsim 20~\GeV$.
A possible explanation is the normalisation of $\rho_c$ to $Q^2$ 
in eq.~(\ref{jetmass}). 
In the Breit current hemisphere multi-parton final states tend to produce 
visible energy $E_{vis} > Q/2$,
which is different from the quark parton model expectation
and from the situation in $e^+e^-$ annihilation.
Indeed, a normalisation to $4\,E^2_{vis}$ instead of $Q^2$
gives better agreement between
$e\,p$ and $e^+e^-$ data, particularly at high values of $Q$.

As argued above,
one does not expect the event shapes distribution
in deep inelastic $e\,p$ scattering to correspond exactly to those
for $e^+e^-$ annihilation.
There are other differences, which are not related to the analysis method. 
The $e^+e^-$ data contain a considerably larger fraction of bottom quarks
with different fragmentation properties than the light quarks,
which may be important close to threshold energies.
Furthermore, there are additional ${\cal O}(\alpha_s)$ processes 
in deep inelastic scattering.
Besides the common final state gluon radiation off quarks, in DIS there are 
contributions from initial state gluon radiation and boson gluon fusion.

\section{QCD Calculations and Power Corrections}
\label{qcdcalcul}

\paragraph{Theoretical Framework}
Any $Q$ or energy dependence of the event shape variables can have the
following origins: 
(i) the logarithmic change of the strong coupling constant 
    $\alpha_s(Q) \propto 1/\ln Q$,
    and 
(ii) power corrections or hadronisation effects,
     which are expected to behave like $1/Q$. 
Recent theoretical developments suggest that $1/Q$ corrections are not
necessarily related to hadronisation, but may instead be a universal soft 
gluon phenomenon associated with the behaviour of the running coupling at 
small momentum scales~\cite{webber,dokshitzer}.
`Universal' means that they could be expressible in terms of a few
non-perturbative parameters with calculable process-dependent 
coefficients.
The event shape data will now be analyzed by applying this approach which
relates hadronisation effects in the final state observables to power
corrections. 

The mean value of any infrared safe event shape variable $\langle F \rangle$ 
at a scale $\mu_R$, taken to be $\mu_R = Q$,
can be written according to ref.~\cite{webber} as
\begin{eqnarray}
       \langle F \rangle & = &
     \langle F \rangle^{{\rm pert}} + 
      \langle F \rangle^{{\rm pow}}\ , 
        \label{fmean} \\[1ex]
      \langle F \rangle^{{\rm pert}} 
        & = & c_1\,\alpha_s(\mu_R) + 
        \left ( c_2 + \frac{\beta_0}{2\,\pi}\,\ln \frac{\mu_R}{Q}\,c_1
        \right ) \alpha^2_s(\mu_R)  \ ,
        \label{fpert} \\[1ex]
      \langle F \rangle^{{\rm pow}} & = & 
      a_F\,\frac{16}{3\,\pi}\,\frac{\mu_I}{\mu_R}\, 
      \ln^p \frac{\mu_R}{\mu_I}
      \left [ \,
      \bar{\alpha}_0(\mu_I) - \alpha_s(\mu_R) 
      - \frac{\beta_0}{2\,\pi}\, 
      \left ( \ln\frac{\mu_R}{\mu_I} + \frac{K}{\beta_0} + 1 \right ) 
      \alpha^2_s(\mu_R) \, \right ] \ ,
        \label{fpow}
\end{eqnarray}
where $\beta_0 = 11 - 2/3\,N_{f}$, $K = 67/6 - \pi^2/2 - 5/9\, N_{f}$
and $N_f = 5$ flavours.

The perturbative part, eq.~(\ref{fpert}), 
represents the second order QCD pre\-diction, where the coefficients 
$c_1$ and $c_2$
are calculated in the $\overline{\mbox{MS}}$ scheme at the scale $\mu_R = Q$.
The power corrections, eq.~(\ref{fpow}) with 
coefficients $a_F$ and $p$ ($p = 0$, except $p = 1$ for $\langle B_c \rangle$) 
depending on the observable $F$, 
contain a free non-perturbative parameter $\bar{\alpha}_0(\mu_I)$ to be 
evaluated at some `infrared matching' scale 
$\Lambda_{QCD} \ll \mu_I \ll \mu_R$,
conventionally chosen to be $\mu_I = 2~\GeV$.
The parameter $\bar{\alpha}_0(\mu_I)$ can be interpreted as an effective strong
coupling constant below the matching scale $\mu_I$, 
calculated in leading order.
The dependence on $\mu_I$ is partially compensated by the $\mu_I$ dependence
of the other terms in eq.~(\ref{fpow}).
The dependence on the renormalisation scale $\mu_R$ should help to compensate
the scale dependence of the perturbative part. 
The power corrections are expected to behave like $1/\mu_R$ 
({\em i.e.} $p = 0$)
for the event shape variables discussed in this paper,
except for the jet broadening which is supposed to have an additional
$\ln (\mu_R/\mu_I)$ term ({\em i.e.} $p = 1$). 
However, the prediction for $\langle B_c \rangle$ is considered to be less
reliable~\cite{webber}.
In this approach the data 
can be directly used to determine in a simultaneous fit 
the power correction parameters $\bar{\alpha}_0(\mu_I)$ and 
the strong coupling constant $\alpha_s(M_Z)$  
without assuming any fragmentation model.

\paragraph{Perturbative Predictions}
In deep inelastic scattering the perturbative part, eq.~(\ref{fpert}),
may be obtained via 
\begin{eqnarray}
  \label{meandef}
  \langle F \rangle ^{{\rm pert}} & = &
  \frac{\int_{0}^{F_{\rm max}} F \, 
    \frac{{\rm d} \sigma}{{\rm d}F} \, dF}
  {\int_{0}^{F_{\rm max}} \frac{{\rm d} \sigma}{{\rm d} F} \, dF} 
  \ = \
  \frac{1}{\sigma_{\rm tot}} 
  \int_{0}^{F_{\rm max}} \! F \, \frac{{\rm d} \sigma}{{\rm d}F} \, dF \, .
\end{eqnarray}
To get $\langle F \rangle ^{{\rm pert}}$ to ${\cal O}(\alpha_s^2)$
the Taylor expansion of eq.~(\ref{meandef}) shows that
the integral in the numerator has to be evaluated in second order QCD,
whereas the total cross section $\sigma_{\rm tot}$
%in the denominator 
needs only be known to first order, because the numerator vanishes,
$F \equiv 0$, in the quark parton model.

The perturbative calculations can be performed with the
MEPJET~\cite{mepjet} and DISENT~\cite{disent} programs.
Both programs treat deep inelastic $e\,p$ scattering to ${\cal O}(\alpha_s^2)$ 
via single $\gamma$ exchange.
The neglect of $Z$ exchange has no influence on event shape parameters in the
range of $Q$ investigated.
The lack of virtual two-loop diagrams, however, leads to the restriction
that $\sigma_{\rm tot}$ can only be calculated to ${\cal O}(\alpha_s)$,
which, as explained above, is sufficient for the present analysis.

MEPJET applies the so-called `phase space slicing method' 
%and the `crossing function' technique 
for the integration.
The real emission of partons
(quarks, antiquarks and gluons) is calculated exactly for a two-parton
resolution parameter $s_{ij} > s_{min}$, the invariant mass squared, 
with typical values of $s_{min} \sim 0.01~\GeV^2$.
Soft and collinear approximations are used in the region where at least one
parton pair has $s_{ij} < s_{min}$.
The final state infrared and collinear divergences cancel against virtual loop
and tree diagrams.
The cut on $s_{min}$ prevents the integration of the event shape distributions
$F \, {\rm d} \sigma/{\rm d}F$ being carried out over the whole phase space. 
Instead, a lower bound $F_{\rm cut} > 0$, which depends on the event shape 
under consideration, has to be imposed. 

DISENT uses the `subtraction method' and
applies dipole factorisation formulae
for the same set of ${\cal O}(\alpha_s^2)$ diagrams.
Since no $s_{min}$ cut is needed and special care has been
taken concerning the numerical integration of $F \, {\rm d} \sigma /{\rm d}F$,
which still contains integrable singularities,
one is allowed to use the complete phase space
$0 \leq F \leq F_{\rm max}$.
Therefore the DISENT results are applied throughout the present analysis. 
In the overlapping phase space regions with $F > F_{\rm cut}$
the differential event shape
distributions of MEPJET and DISENT agree very well
and the mean values from both programs yield compatible results
to within the statistical accuracy,
which is less than $1\%$ in first order 
and about $3\%$ in second order perturbative QCD.

\begin{figure}[p] \centering \unitlength 1mm
  \begin{picture}(170,200)
    \put(-5,0)  
     {\mbox{
       \epsfig{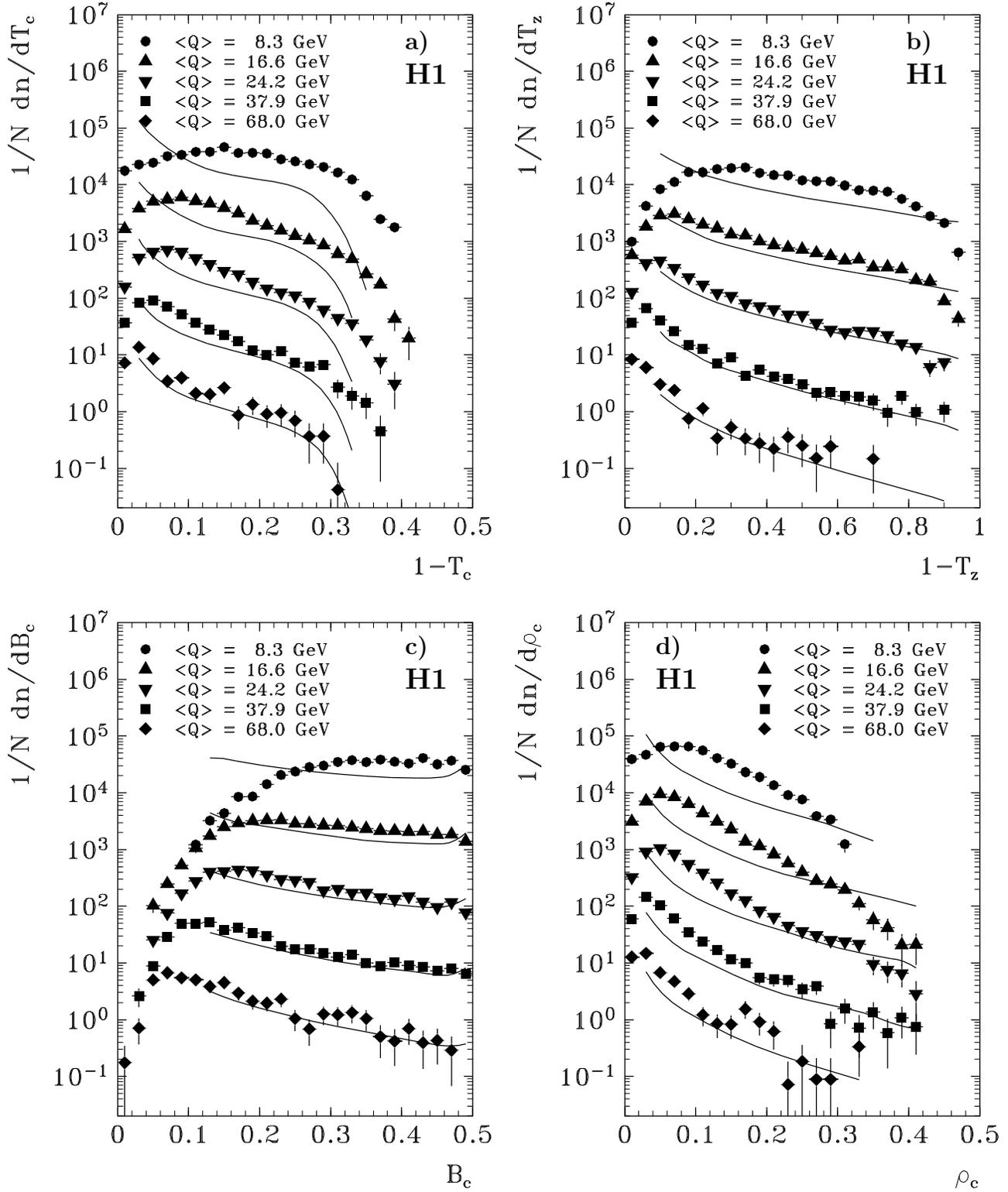} } } 
    \put( 67,198.5) {\bf a) }
    \put(153,198.5) {\bf b) }
    \put( 67, 94.5) {\bf c) }
    \put(110, 94.5) {\bf d) }
    \put( 67,192.5) {\large\bf H1 }
    \put(153,192.5) {\large\bf H1 }
    \put( 67, 88.5) {\large\bf H1 }
    \put(110, 88.5) {\large\bf H1 }
  \end{picture}
  \caption{Differential event shape distributions of
           {\bf a)} thrust $1/N\,dn/d(1 - T_c)$,
           {\bf b)} thrust $1/N\,dn/d(1 - T_z)$,
           {\bf c)} jet broadening $1/N\,dn/dB_c$ and
           {\bf d)} jet mass $1/N\,dn/d\rho_c$.
           H1~DIS $e\,p$ data (full symbols, only statistical errors shown)
           are compared with second order QCD calculations (---).
           The spectra for $\langle Q \rangle = 8.3 - 68~\GeV$
           are multiplied by factors of $10^n$ ($n = 0, 4$)}
  \label{wdistr}
\end{figure}

Fig.~\ref{wdistr} shows the differential event shape spectra at the parton
level based on the second order QCD calculations
in comparison with the data.
The lepton and partons are selected in the same way as in 
the data by requiring the criteria 1, 2, 6 and 7 of
section~\ref{analysis}.
One clearly sees the strong impact of hadronisation effects,
particularly at low values of $Q$.
Towards higher values of $Q$ the hadronisation becomes less important and
the data and parton distributions approach each other.
It is a consequence of the concept of ref.~\cite{webber}
that the mean values of the event shape spectra of the observed data 
and of the partons can be related
by a single power correction parameter $\bar{\alpha}_0$.

For each event shape variable
the coefficients $c_1$ and $c_2$ are obtained from a
fit of eq.~(\ref{fpert}) to the $Q$ dependence of the mean values,
which themselves are calculated
with DISENT from  eq.~(\ref{meandef}).
The theoretical calculations are performed by using the MRS~H parton 
distributions~\cite{mrsh} and the corresponding value of the strong coupling
constant.
The coefficients $c_1$, $c_2$ of the perturbative part, eq.~(\ref{fpert}),
and the coefficient $a_F$~\cite{webber}
of the power corrections, eq.~(\ref{fpow}), which are
used in the following QCD analysis of the event shape variables
are given in table~\ref{cparameters}.

\paragraph{QCD Fits}
Fits to the data of fig.~\ref{meandata} and table~\ref{means} 
show that contributions from second order power corrections
$\propto 1/Q^{2}$ are not required by the data.
Such terms may occur because of higher-twist corrections and
would be sensitive to the low $Q$ measurements.
All event shape variables can be well described by first order power
corrections $\propto 1/Q$ and an exponent $p = 0$ for the logarithmic
term. 

Note that the power corrections for the jet broadening are expected to have
a $Q$ dependence different from the other event shapes variables.
However, the ansatz 
$\langle B_c \rangle^{{\rm pow}} \propto 1/\mu_R\,\ln (\mu_R/\mu_I)$
of  eq.~(\ref{fpow}) cannot be supported by the data.
%The additional logarithmic term 
%%varies by a factor of $\sim 2.7$ over the measured $Q$ range and it 
%essentially counteracts the
%$\alpha_s(\mu_R) \propto 1/\ln(\mu_R/\Lambda)$ dependence of the strong 
%coupling and gives an unacceptable fit.
The additional logarithmic term varies by a factor of $\sim 2.7$ and
leads to a less steep $Q$ dependence, resulting in an unacceptable fit.
%An acceptable fit can only be accommodated if the strong coupling constant is
%allowed to approach unreasonably low values, {\em i.e.} 
%$\alpha_s(M_Z) \rightarrow 0$.
%At the same time the power correction parameter $\bar{\alpha}_0$ decreases 
%by $\sim 25\,\%$.
A modification of the logarithmic term
to $\ln (\mu_R/\mu_0)$ with $\mu_0$ treated as a free parameter
also does not give a satisfactory solution.
Therefore this logarithmic term is discarded in the following analysis.
This observation disagrees with the conjectured $Q$ behaviour of
ref.~\cite{webber}. 

\begin{figure}[p] \centering \unitlength 1mm
  \begin{picture}(170,200)
    \put(-5,0)  
     {\mbox{
       \epsfig{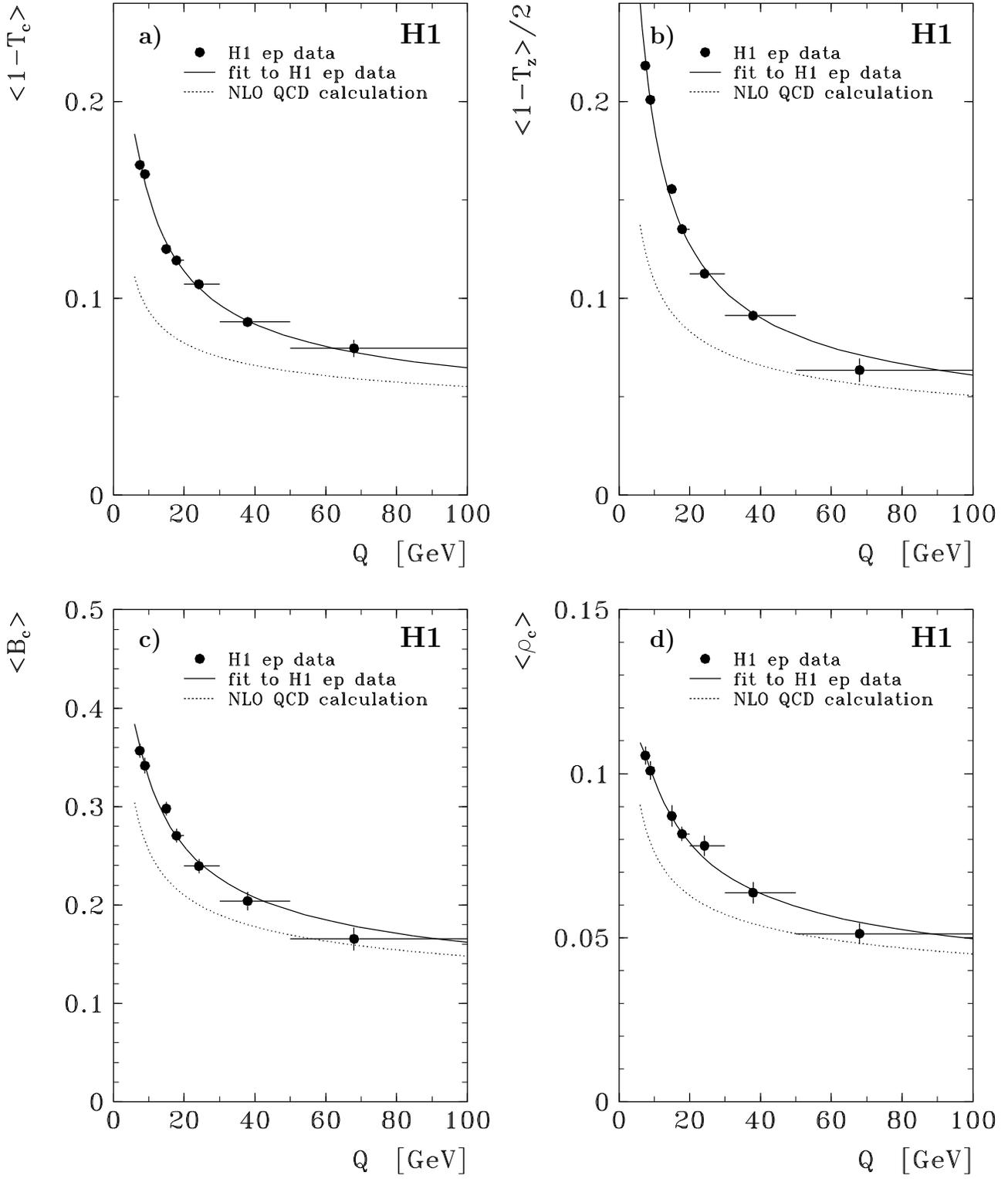} } } 
    \put( 22,197.5) {\bf a) }
    \put(110,197.5) {\bf b) }
    \put( 22, 93.5) {\bf c) }
    \put(110, 93.5) {\bf d) }
    \put( 67,197.5) {\large\bf H1 }
    \put(155,197.5) {\large\bf H1 }
    \put( 67, 93.5) {\large\bf H1 }
    \put(155, 93.5) {\large\bf H1 }
  \end{picture}
  \caption{Mean event shape variables as a function of Q for
           {\bf a)} $\langle 1 - T_c \rangle$,
           {\bf b)} $\langle 1 - T_z \rangle / 2$,
           {\bf c)} $\langle B_c \rangle$, and
           {\bf d)} $\langle \rho_c \rangle$.
           H1 DIS $e\,p$ data ($\bullet$, errors include statistics
           and systematics) are compared with QCD fits (---)
           and second order QCD calculations ($\cdot\cdot\cdot$)}
  \label{fits}
\end{figure}

The results of the QCD analysis, assuming $\mu_R = Q$ and $\mu_I = 2~\GeV$,
are shown in fig.~\ref{fits}. 
The data are well represented by the fits to 
eqs.~(\ref{fmean}) -- (\ref{fpow}). 
The second order perturbative QCD predictions are also shown. 
The power corrections, or hadronisation contributions, are substantial at low
values of $Q$, but become less important with increasing energy.
A comparison with the corresponding LEPTO curves of 
fig.~\ref{meandata} shows that this event generator 
adequately describes the experimental data. 
However, the LEPTO parton level distributions look very different to the 
DISENT or MEPJET calculations.

Theoretical uncertainties in the determination of $\alpha_s(M_Z)$ and
$\bar{\alpha}_0$ come from the accuracy of the QCD calculations and 
from the choice of different possible scales.
Varying the anti-correlated coefficients $c_1$ and $c_2$ of eq.~(\ref{fpert})
within their errors leads to typical changes of
$\delta\alpha_s^{calc} = \pm 0.001$ and 
$\delta\bar{\alpha}_0^{calc} = \pm 0.002$.
Using a different parton density parametrisation would give negligible
changes to $\alpha_s$ and $\bar{\alpha}_0$.
The separation between the non-perturbative and perturbative contributions is
characterized by the infrared matching scale $\mu_I$ and 
the renormalisation scale $\mu_R$, 
which should satisfy the relation $\Lambda \ll \mu_I \ll \mu_R$.
Requiring the criterion $\mu_R/\mu_I > 3$ implies 
$\mu_I < 2.5~\GeV$ if $\mu_R = Q$
and $\mu_R > 0.8\,Q$ if $\mu_I = 2~\GeV$. 
The upper value of the renormalisation scale is chosen such that the fitted
$\chi^2$ varies by about
the same amount as for the statistical error determination,
typically $\mu_R < 1.5\,Q$.
A variation of the renormalisation scale within $0.8\,Q < \mu_R < 1.5\,Q$ 
results in shifts of the order of 
$\delta\alpha_s^{\mu_R} = \ ^{+0.005}_{-0.004}$ and 
$\delta\bar{\alpha}_0^{\mu_R} = \pm 0.05$.
The range of the infrared matching scale is chosen to be symmetric,
$\mu_I = 2.0 \pm 0.5~\GeV$, which leads to typical changes of
$\delta\alpha_s^{\mu_I} = \pm 0.002$, while the power corrections are
proportional to $\mu_I$.
All the theoretical uncertainties are added in quadrature.

\section{Results of QCD Analysis}

The final results for the power correction parameters $\bar{\alpha}_0$ and
the strong coupling constant $\alpha_s(M_Z)$
in the $\overline{\mbox{MS}}$ scheme
are compiled in table~\ref{fitresults}.
Note that the overall errors are dominated by systematic effects associated 
with the theory.
The largest contribution comes from the renormalisation scale dependence, 
which could be reduced once ${\cal O}(\alpha_s^3)$ calculations become
available. 
However, there remains the interplay between the non-perturbative and
perturbative regions. 
An effective means to get a larger separation between the renormalisation
and the infrared matching scales would be
to select data at higher values of $Q$ at the expense of a lower 
sensitivity to both the power correction parameter and
the strong coupling constant.

All event shape analyses give results consistent with each other  
for $\bar{\alpha}_0$ and $\alpha_s(M_Z)$.
It should be noted that the fitted parameters $\bar{\alpha}_0$ are 
correlated with the preferred value of $\alpha_s(M_Z)$. The correlation
coefficients change sign and magnitude for the various event shape variables.

It is particularly remarkable that the power correction parameters 
$\bar{\alpha}_0$ of the event shapes under study are all of the same size 
to within $\pm 15\%$.
The individual values are consistent with being universal  
and clearly support the concept and computations of ref.~\cite{webber}
of power corrections to the mean hadronic event shapes  
in deep inelastic $e\,p$ scattering.
It is notable, however, 
that the power correction parameter for the jet broadening is
slightly lower than those of the other event shapes, which may be an
indication that the adopted parametrisation is not completely correct,
as noted in section~\ref{qcdcalcul}.

The $\alpha_s(M_Z)$ values show a larger spread than expected from the
experimental errors alone. They become compatible with each other only if the
theoretical uncertainties are taken into account.
This indicates the importance of the influence 
of higher order QCD corrections,
which may affect each event shape variable differently.
A similar situation was observed in ${\cal O}(\alpha_s^2)$ event shape
analyses performed by $e^+e^-$ experiments at LEP, 
see {\em e.g.} discussion in ref.~\cite{bethke}.

An important result is the $1/Q$ dependence of the power corrections to the
mean event shape variables. 
The similarity of the $\bar{\alpha}_0$ values suggests that
the assumption of universal power corrections may be further tested
by performing a common two-parameter fit to both thrust and 
the jet mass data, each of which exploits a different event shape pro\-perty.
The jet broadening data are neglected because of the arguments given above.
%Possible correlations between the event shape variables are not accounted for,
%but the correlation between $\bar{\alpha}_0$ and $\alpha_s(M_Z)$ is much
%reduced. 
In the common fit no account is taken for
possible correlations between the event shape variables.
The correlation between $\bar{\alpha}_0$ and $\alpha_s(M_Z)$ is much
less than in the individual fits. 
The common fit (see table~\ref{fitresults})
gives a relatively large $\chi^2/{\rm ndf} \simeq 2$,
a possible reason being the missing higher order QCD corrections.
The results of the fit are
$\bar{\alpha}_0 = 0.491 \pm 0.003~({\mbox{exp}}) 
  \ ^{+0.079}_{-0.042}~({\mbox{theory}})$ 
for the power correction parameter and 
$\alpha_s(M_Z) = 0.118 \pm 0.001~({\mbox{exp}})
  \ ^{+0.007}_{-0.006}~({\mbox{theory}})$
for the strong coupling constant in the $\overline{\mbox{MS}}$ scheme, 
where the experimental errors are determined from the $\chi^2$ contour.
It is clear from the sizes of the experimental and theoretical uncertainties
that a better understanding of higher order QCD corrections to the event shape
variables is necessary for further improvements.

From fits to the energy dependence ($Q = \sqrt{s}$)
of thrust $\langle 1 - T_{ee} \rangle$
and the heavy jet mass $\langle M^2_H/s \rangle$ of 
$e^+e^-$ data~\cite{eedata}, 
and using the the same approach with the QCD prescriptions 
of ref.~\cite{dokshitzer},
one obtains values for the power correction parameters which are consistent 
with the $e\,p$ analyses, see table~\ref{fitresults}.
In a similar analysis with a different choice of $e^+e^-$ experiments  
DELPHI~\cite{eedata} finds consistent results,
but a slightly smaller value of $\alpha_s(M_Z)$ for the thrust variable.
These observations suggest that the power correction parameters
$\bar{\alpha}_0$ are universal to 
within $\sim 20\% $ in both deep inelastic  
$e\,p$ scattering and $e^+e^-$ annihilation.\footnote{Notice that
  any analysis of the mean event shape variables
  in $e^+e^-$ experiments also suffers from
  missing higher order QCD corrections.}

\section{Conclusion} 

The event shape variables thrust, jet broadening and jet mass are studied
for the first time in the current hemisphere of the Breit frame 
in deep inelastic $e\,p$ scattering.
Differential distributions and mean values are measured over a wide 
range of momentum transfers $Q$ from $7~\GeV$ to $100~\GeV$. 
The mean values of the event shapes exhibit a strong $Q$ dependence. 
They decrease with rising $Q$, {\em i.e.} the energy flow in the Breit current
hemisphere becomes more collimated.  
The means of thrust and jet mass show a similar 
energy dependence to that measured in 
$e^+e^-$ annihilation at the scale $Q = \sqrt{s_{ee}}$.
Differences can be plausibly attributed to the different QCD dynamics for,
and to the analysis methods which are applied to,
$e\,p$ and $e^+e^-$ scattering.

Using ${\cal O}(\alpha_s^2)$ QCD calculations and a specific model to describe
the influence of non-perturbative effects on hadronic final states,
the $Q$ dependences of the 
mean event shape parameters are fitted to determine simultaneously the power
or hadronisation corrections and the strong coupling constant 
$\alpha_s(M_Z)$ in the $\overline{\mbox{MS}}$ scheme 
without assuming any fragmentation model.
The power corrections decrease with the expected power of $1/Q$ and can be
described by a common universal parameter $\bar{\alpha}_0$ which is
consistent with the values found in $e^+e^-$ experiments. 
However, the conjectured $1/Q\,\ln Q$ dependence of the jet broadening power
corrections cannot be supported by the $e\,p$ data.
The present precision of both the power correction para\-meters and the
strong coupling constant
is limited by theoretical uncertainties due to unknown 
higher order QCD corrections, which are to some extent accounted for by 
a variation of the renormalisation and infrared matching scales. 

%
%The results for the strong coupling constant
%may be compared with the average value of 
%$\alpha_s(M_Z) = 0.122 \pm 0.007$ obtained from a large set of 
%event shape analyses in $e^+e^-$ experiments at the $Z$ resonance~\cite{pdg}, 
%where the error is totally dominated by theoretical uncertainties associated 
%with the choice of renormalisation scale and the hadronisation effects of the 
%Monte Carlo generators used to fit the various quantities. 

\paragraph{Acknowledgments} 
We are grateful to the HERA machine group whose outstanding 
efforts have made and continue to make this experiment possible. We thank 
the engineers and technicians for their work in constructing and now 
maintaining the H1 detector, our funding agencies for financial support, the 
DESY technical staff for continual assistance, and the DESY directorate for the
hospitality which they extend to the non--DESY members of the collaboration. 
We gratefully acknowledge valuable discussions with E.~Mirkes, M.~Seymour and 
B.R.~Webber.

\clearpage

\begin{table}[htb]
\begin{center}
    \begin{tabular}{c c}
      \hdick \\[-1.5ex]
      $\langle Q \rangle~[\GeV]$ \ &  
      $\langle 1 - T_c \rangle$    
      \\[1ex] \hdick \\[-1.5ex]
      7.47 & $ 0.1678 \pm 0.0019 \pm 0.0023 $ 
      \\[.4ex]
      8.91 & $ 0.1631 \pm 0.0021 \pm 0.0019 $ 
      \\[.4ex]
      14.9 & $ 0.1251 \pm 0.0016 \pm 0.0021 $ 
      \\[.4ex]
      17.8 & $ 0.1193 \pm 0.0012 \pm 0.0022 $ 
      \\[.4ex]
      24.2 & $ 0.1072 \pm 0.0012 \pm 0.0025 $ 
      \\[.4ex]
      37.9 & $ 0.0880 \pm 0.0018 \pm 0.0021 $ 
      \\[.4ex]
      68.0 & $ 0.0746 \pm 0.0037 \pm 0.0023 $ 
      \\[1ex] \hdick 
    \end{tabular}
    \hspace{1.5cm}
    \begin{tabular}{c c}
      \hdick \\[-1.5ex]
      $\langle Q \rangle~[\GeV]$ \ &  
      $\langle 1 - T_z \rangle \, / \, 2$    
      \\[1ex] \hdick \\[-1.5ex]
      7.47 & $ 0.2182 \pm 0.0019 \pm 0.0024 $ 
      \\[.4ex]
      8.91 & $ 0.2009 \pm 0.0022 \pm 0.0023 $ 
      \\[.4ex]
      14.9 & $ 0.1555 \pm 0.0018 \pm 0.0023 $ 
      \\[.4ex]
      17.8 & $ 0.1351 \pm 0.0014 \pm 0.0025 $ 
      \\[.4ex]
      24.2 & $ 0.1125 \pm 0.0013 \pm 0.0024 $ 
      \\[.4ex]
      37.9 & $ 0.0912 \pm 0.0019 \pm 0.0023 $ 
      \\[.4ex]
      68.0 & $ 0.0635 \pm 0.0035 \pm 0.0049 $ 
      \\[1ex] \hdick 
    \end{tabular}
    \\ \vspace{1.5cm}
    \begin{tabular}{c c}
      \hdick \\[-1.5ex]
      $\langle Q \rangle~[\GeV]$ \ &  
      $\langle B_c \rangle$    
      \\[1ex] \hdick \\[-1.5ex]
      7.47 & $ 0.3566 \pm 0.0019 \pm 0.0071 $ 
      \\[.4ex]
      8.91 & $ 0.3414 \pm 0.0023 \pm 0.0076 $ 
      \\[.4ex]
      14.9 & $ 0.2978 \pm 0.0018 \pm 0.0069 $ 
      \\[.4ex]
      17.8 & $ 0.2704 \pm 0.0013 \pm 0.0067 $ 
      \\[.4ex]
      24.2 & $ 0.2394 \pm 0.0014 \pm 0.0070 $ 
      \\[.4ex]
      37.9 & $ 0.2039 \pm 0.0023 \pm 0.0093 $ 
      \\[.4ex]
      68.0 & $ 0.1654 \pm 0.0046 \pm 0.0107 $ 
      \\[1ex] \hdick 
    \end{tabular}
   \hspace{1.5cm}
    \begin{tabular}{c c}
      \hdick \\[-1.5ex]
      $\langle Q \rangle~[\GeV]$ \ &  
      $\langle \rho_c \rangle$    
      \\[1ex] \hdick \\[-1.5ex]
      7.47 & $ 0.1055 \pm 0.0015 \pm 0.0023 $ 
      \\[.4ex]
      8.91 & $ 0.1009 \pm 0.0015 \pm 0.0024 $ 
      \\[.4ex]
      14.9 & $ 0.0871 \pm 0.0010 \pm 0.0031 $ 
      \\[.4ex]
      17.8 & $ 0.0816 \pm 0.0008 \pm 0.0020 $ 
      \\[.4ex]
      24.2 & $ 0.0780 \pm 0.0008 \pm 0.0030 $ 
      \\[.4ex]
      37.9 & $ 0.0637 \pm 0.0013 \pm 0.0030 $ 
      \\[.4ex]
      68.0 & $ 0.0512 \pm 0.0024 \pm 0.0021 $ 
      \\[1ex] \hdick %\\[-1.5ex]
    \end{tabular}
\end{center}
\caption{Mean values of the event shape variables
      thrust
      $\langle 1 - T_c \rangle$ and    
      $\langle 1 - T_z \rangle / 2$, 
      jet broadening
      $\langle B_c \rangle$ and
      jet mass
      $\langle \rho_c \rangle$      
      as a function of $Q$.
      The first error is statistical, the second systematic}
\label{means}
\end{table}

\begin{table}[htb]
\begin{center}
    \begin{tabular}{l c c c}
      \hdick \\[-1.5ex]
      Observable & $c_1$ & $c_2$ & $a_F$ 
      \\[1ex] \hdick \\[-1.5ex]
      $\langle 1 - T_c \rangle$ &
        $ \ 0.384 \pm 0.033 \ $ &
        $ \ 0.57  \pm 0.21 \ $ 
        & $ 1 $
      \\[.4ex]  %\hline
      $\langle 1 - T_z \rangle \, / \, 2$ &
        $ \ 0.053 \pm 0.033 \ $ &
        $ \  3.45 \pm 0.23 \ $ 
        & $ 1 $
      \\[.4ex]  %\hline
      $\langle B_c \rangle$ &
        $ \ 0.990 \pm 0.121 \ $ &
        $ \ 2.39  \pm 0.86 \ $ 
        & $ 2 $
      \\[.4ex]  %\hline
      $\langle \rho_c \rangle$ &
        $ \ 0.359 \pm 0.048 \ $ &
        $  -0.05 \pm 0.30 \ $ 
        & $ 1/2 $
      \\[1ex] \hdick % \\[-1.5ex]
    \end{tabular}
\end{center}
\caption{Coefficients $c_1$, $c_2$ and $a_F$ 
  of the event shape variables used in the QCD fits}
\label{cparameters}
\end{table}

\clearpage
\begin{table}[htb]
\begin{center}
    \begin{tabular}{l c c c}
      \hdick \\[-1.5ex]
      Observable     &
        $\bar{\alpha}_0(\mu_I = 2~\GeV)$ & $\alpha_s(M_Z)$ &
        \ $\chi^2/\mbox{ndf}$ \
      \\[1ex] \hdick \\[-1.5ex]
      H1 $e\,p$ data & \\[.6ex] 
      $\langle 1 - T_c \rangle$ &
        $ \ 0.497 \pm 0.005 \ ^{+0.070}_{-0.036} \ $ &
        $ \ 0.123 \pm 0.002 \ ^{+0.007}_{-0.005} \ $ &
        $ \ 5.0/5 \ $ 
      \\[.4ex]  
      $\langle 1 - T_z \rangle \, / \, 2$ &
        $ \ 0.507 \pm 0.008 \ ^{+0.109}_{-0.051} \ $ &
        $ \ 0.115 \pm 0.002 \ ^{+0.007}_{-0.005} \ $ &
        $ \ 8.5/5 \ $ 
      \\[.4ex]  
      $\langle B_c \rangle$ &
        $ \ 0.408 \pm 0.006 \ ^{+0.036}_{-0.022} \ $ & 
        $ \ 0.119 \pm 0.003 \ ^{+0.007}_{-0.004} \ $ &
        $ \ 5.3/5 \ $ 
      \\[.4ex]  
      $\langle \rho_c \rangle$ &
        $ \ 0.519 \pm 0.009 \ ^{+0.025}_{-0.020} \ $ &
        $ \ 0.130 \pm 0.003 \ ^{+0.007}_{-0.005} \ $ &
        $ \ 3.1/5 \ $ 
      \\[1ex]
      \hdick \\[-1.5ex]
       common fit & \\
      $T_c, \ T_z, \ \rho_c$ & 
        $ \ 0.491 \pm 0.003 \ ^{+0.079}_{-0.042} \ $ &
        $ \ 0.118 \pm 0.001 \ ^{+0.007}_{-0.006} \ $ &
        $ \ 39/19 \ $ 
     \\[1ex] \hdick \\[-1.5ex]
      $e^+e^-$ data \\[.6ex] 
      $\langle 1 - T_{ee} \rangle$ &
        $ \ 0.519 \pm 0.009 \ ^{+0.093}_{-0.039} \ $ &
        $ \ 0.123 \pm 0.001 \ ^{+0.007}_{-0.004} \ $ &
        $ \ 10.9/14 \ $ 
      \\[.4ex]  
      $\langle M^2_H/s \rangle$ &
        $ \ 0.431 \pm 0.020 \ ^{+0.071}_{-0.030} \ $ &
        $ \ 0.115 \pm 0.002 \ ^{+0.005}_{-0.003} \ $ &
        $ \ 7.8/14 \ $ 
       \\[1ex] \hdick % \\[-1.5ex]
  \end{tabular}
\end{center}
\caption{Results on power correction parameters $\bar{\alpha}_0$
  and the strong coupling constant $\alpha_s(M_Z)$ 
  from fits to the $Q$ dependence of the mean event shape variables.
  In the common fit possible correlations between the event shape variables
  are not taken into account.
  The first error is experimental, the second error represents theoretical 
  uncertainties. 
  The $e^+e^-$ experiments~\protect\cite{eedata} use the heavy jet mass $M_H$ }
\label{fitresults}
\end{table}

\clearpage

\begin{figure}[p] \centering \unitlength 1mm
  \begin{picture}(170,200)
    \put(-5,0)  
     {\mbox{
       \epsfig{file=cdistr.eps,%
         bbllx=0pt,bblly=0pt,bburx=500pt,bbury=600pt,clip=,%
         angle=0,width=17.5cm} } } 
    \put( 67,198.5) {\bf a) }
    \put(153,198.5) {\bf b) }
    \put( 67, 94.5) {\bf c) }
    \put(110, 94.5) {\bf d) }
    \put( 67,192.5) {\large\bf H1 }
    \put(153,192.5) {\large\bf H1 }
    \put( 67, 88.5) {\large\bf H1 }
    \put(110, 88.5) {\large\bf H1 }
  \end{picture}
  \caption{Differential event shape distributions of
           {\bf a)} thrust $1/N\,dn/d(1 - T_c)$,
           {\bf b)} thrust $1/N\,dn/d(1 - T_z)$,
           {\bf c)} jet broadening $1/N\,dn/dB_c$ and
           {\bf d)} jet mass $1/N\,dn/d\rho_c$.
           H1 DIS $e\,p$ data (full symbols, only statistical errors shown)
           are compared with LEPTO Monte Carlo simu\-lations (---).
           The spectra for $\langle Q \rangle = 8.3 - 68~\GeV$
           are multiplied by factors of $10^n$ ($n = 0, 4$)}
  \label{cdistr}
\end{figure}

\end{document}

%% file: h1auts.tex
%   H1AUTS  Author list by names, no. of authors  379
%           status: 26/03/97   09.19.14
 C.~Adloff$^{35}$,                %WUPP-ST                  Adloff              
 S.~Aid$^{13}$,                   %HAM2-LEFT    8/96        Aid                 
 M.~Anderson$^{23}$,              %MANC-ST  10/95           Anderson            
 V.~Andreev$^{26}$,               %LPI -PD                  Andreev             
 B.~Andrieu$^{29}$,               %ECPL-PD                  Andrieu             
 V.~Arkadov$^{36}$,               %ZEUT-ST    10/96         Arkadov             
 C.~Arndt$^{11}$,                 %DESY-ST   1/96           Arndt               
 I.~Ayyaz$^{30}$,                 %PARI-ST       5/96       Ayyaz               
 A.~Babaev$^{25}$,                %ITEP-PD                  Babaev              
 J.~B\"ahr$^{36}$
% ...

%% file: h1inst.tex
%     H1 Institutes as appearing on publications
 $ ^1$ I. Physikalisches Institut der RWTH, Aachen, Germany$^ a$ \\
 $ ^2$ III. Physikalisches Institut der RWTH, Aachen, Germany$^ a$ \\
 $ ^3$ School of Physics and Space Research, University of Birmingham,
                             Birmingham, UK$^ b$\\
 $ ^4$ Inter-University Institute for High Energies ULB-VUB, Brussels;
   Universitaire Instelling Antwerpen, Wilrijk; Belgium$^ c$ \\
 $ ^5$ Rutherford Appleton Laboratory, Chilton, Didcot, UK$^ b$ \\
 $ ^6$ Institute for Nuclear Physics, Cracow, Poland$^ d$  \\
 $ ^7$ Physics Department and IIRPA,
         University of California, Davis, California, USA$^ e$ \\
 $ ^8$ Institut f\"ur Physik, Universit\"at Dortmund, Dortmund,
                                                  Germany$^ a$\\
 $ ^{9}$ DSM/DAPNIA, CEA/Saclay, Gif-sur-Yvette, France \\
 $ ^{10}$ Department of Physics and Astronomy, University of Glasgow,
                                      Glasgow, UK$^ b$ \\
 $ ^{11}$ DESY, Hamburg, Germany$^a$ \\
 $ ^{12}$ I. Institut f\"ur Experimentalphysik, Universit\"at Hamburg,
                                     Hamburg, Germany$^ a$  \\
 $ ^{13}$ II. Institut f\"ur Experimentalphysik, Universit\"at Hamburg,
                                     Hamburg, Germany$^ a$  \\
 $ ^{14}$ Max-Planck-Institut f\"ur Kernphysik,
                                     Heidelberg, Germany$^ a$ \\
 $ ^{15}$ Physikalisches Institut, Universit\"at Heidelberg,
                                     Heidelberg, Germany$^ a$ \\
 $ ^{16}$ Institut f\"ur Hochenergiephysik, Universit\"at Heidelberg,
                                     Heidelberg, Germany$^ a$ \\
 $ ^{17}$ Institut f\"ur Reine und Angewandte Kernphysik, Universit\"at
                                   Kiel, Kiel, Germany$^ a$\\
 $ ^{18}$ Institute of Experimental Physics, Slovak Academy of
                Sciences, Ko\v{s}ice, Slovak Republic$^{f,j}$\\
 $ ^{19}$ School of Physics and Chemistry, University of Lancaster,
                              Lancaster, UK$^ b$ \\
 $ ^{20}$ Department of Physics, University of Liverpool,
                                              Liverpool, UK$^ b$ \\
 $ ^{21}$ Queen Mary and Westfield College, London, UK$^ b$ \\
 $ ^{22}$ Physics Department, University of Lund,
                                               Lund, Sweden$^ g$ \\
 $ ^{23}$ Physics Department, University of Manchester,
                                          Manchester, UK$^ b$\\
 $ ^{24}$ CPPM, Universit\'{e} d'Aix-Marseille II,
                          IN2P3-CNRS, Marseille, France\\
 $ ^{25}$ Institute for Theoretical and Experimental Physics,
                                                 Moscow, Russia \\
 $ ^{26}$ Lebedev Physical Institute, Moscow, Russia$^ f$ \\
 $ ^{27}$ Max-Planck-Institut f\"ur Physik,
                                            M\"unchen, Germany$^ a$\\
 $ ^{28}$ LAL, Universit\'{e} de Paris-Sud, IN2P3-CNRS,
                            Orsay, France\\
 $ ^{29}$ LPNHE, Ecole Polytechnique, IN2P3-CNRS,
                             Palaiseau, France \\
 $ ^{30}$ LPNHE, Universit\'{e}s Paris VI and VII, IN2P3-CNRS,
                              Paris, France \\
 $ ^{31}$ Institute of  Physics, Czech Academy of
                    Sciences, Praha, Czech Republic$^{f,h}$ \\
 $ ^{32}$ Nuclear Center, Charles University,
                    Praha, Czech Republic$^{f,h}$ \\
 $ ^{33}$ INFN Roma~1 and Dipartimento di Fisica,
               Universit\`a Roma~3, Roma, Italy   \\
 $ ^{34}$ Paul Scherrer Institut, Villigen, Switzerland \\
 $ ^{35}$ Fachbereich Physik, Bergische Universit\"at Gesamthochschule
               Wuppertal, Wuppertal, Germany$^ a$ \\
 $ ^{36}$ DESY, Institut f\"ur Hochenergiephysik,
                              Zeuthen, Germany$^ a$\\
 $ ^{37}$ Institut f\"ur Teilchenphysik,
          ETH, Z\"urich, Switzerland$^ i$\\
 $ ^{38}$ Physik-Institut der Universit\"at Z\"urich,
                              Z\"urich, Switzerland$^ i$ \\
\smallskip
 $ ^{39}$ Institut f\"ur Physik, Humboldt-Universit\"at,
               Berlin, Germany$^ a$ \\
 $ ^{40}$ Rechenzentrum, Bergische Universit\"at Gesamthochschule
               Wuppertal, Wuppertal, Germany$^ a$ \\
% $ ^{41}$ Visitor from Physics Dept. University Louisville, USA \\
 
%\smallskip
% $ ^{\dagger}$ Deceased \\
 
\bigskip
 $ ^a$ Supported by the Bundesministerium f\"ur Bildung, Wissenschaft,
        Forschung und Technologie, FRG,
        under contract numbers 6AC17P, 6AC47P, 6DO57I, 6HH17P, 6HH27I,
        6HD17I, 6HD27I, 6KI17P, 6MP17I, and 6WT87P \\
 $ ^b$ Supported by the UK Particle Physics and Astronomy Research
       Council, and formerly by the UK Science and Engineering Research
       Council \\
 $ ^c$ Supported by FNRS-NFWO, IISN-IIKW \\
 $ ^d$ Partially supported by the Polish State Committee for Scientific 
       Research, grant no. 115/E-343/SPUB/P03/120/96 \\
 $ ^e$ Supported in part by USDOE grant DE~F603~91ER40674 \\
 $ ^f$ Supported by the Deutsche Forschungsgemeinschaft \\
 $ ^g$ Supported by the Swedish Natural Science Research Council \\
 $ ^h$ Supported by GA \v{C}R  grant no. 202/96/0214,
       GA AV \v{C}R  grant no. A1010619 and GA UK  grant no. 177 \\
 $ ^i$ Supported by the Swiss National Science Foundation \\
 $ ^j$ Supported by VEGA SR grant no. 2/1325/96 \\

%% file: paper.bbl
\clearpage

\begin{thebibliography}{99}
\bibitem{bethke} S.~Bethke, Proceedings {\em QCD~94}, Montpellier,
                 ed. S~Narison, Nucl. Phys. B (Proc. Suppl.) 39~B, C~(1995),
                 p. 198.
\bibitem{mepjet} E.~Mirkes and D.~Zeppenfeld,
                 Phys. Lett. B 380 (1996) 205 and hep-ph/9511448; 
                 E.~Mirkes, private communication.
\bibitem{disent} S.~Catani and M.~Seymour,
                 % Phys. Lett. B 378 (1996) 287 and hep-ph/9602277;
                 Nucl. Phys. B 485 (1997) 291 and hep-ph/9605323;
                 M.~Seymour, private communication.
\bibitem{webber} B.R.~Webber, Proceedings  
                 {\em Workshop on Deep Inelastic Scattering
                 and QCD}, Paris (1995), eds. J.F.~Laporte and Y.~Sirois,
                 p. 115; 
                 B.R.~Webber, private communication;
                 \\ M.~Dasgupta and B.R.~Webber, 
                 preprint Cavendish-HEP-96/5 and hep-ph/9704297.
\bibitem{h1det} H1 Collaboration, I.~Abt et al.,
%                {\em The H1 Detector at HERA}, 
                Nucl. Instr. and Meth. A 386 (1997) 310 and
                Nucl. Instr. and Meth. A 386 (1997) 348.
\bibitem{spacal} H1 SpaCal Group, R.D.~Appuhn et al.,
                 DESY report 96-171 (1996),
                 to be published in Nucl. Instr. and Meth. A.
\bibitem{lepto} G.~Ingelman, Proceedings 
               {\em Physics at HERA}, Hamburg (1991),
               eds. W.~Buch\-m\"uller and G.~Ingelman, vol.~3, p.~1366.
\bibitem{cdm} L.~L\"onnblad, Computer Phys. Comm. 71 (1992) 15. 
\bibitem{mrsh} A.D.~Martin, W.J.~Stirling and R.G.~Roberts,
               Proceedings {\em Workshop on Quantum Field Theory and
                 Theoretical Aspects of High Energy Physics} (1993),
               eds. B.~Geyer and E.M.~Ilgenfritz, p.~11.
\bibitem{herwig} G.~Marchesini et al., Computer Phys. Comm. 67 (1992) 465.
\bibitem{django} G.A.~Schuler and H.~Spiesberger, Proceedings 
                {\em Physics at HERA}, Hamburg (1991),
                eds. W.~Buch\-m\"uller and G.~Ingelman, vol.~3, p.~1419.
\bibitem{eedata} PLUTO Collaboration, Ch.~Berger et al.,
                 Z. Phys. C 12 (1982) 297; \\
                 Mark~II Collaboration, A.~Peterson et al.,
                 Phys. Rev. D 37 (1988) 1; \\
                 TASSO Collaboration, W.~Braunschweig et al., 
                 Z. Phys. C 45 (1989) 11 and
                 Z. Phys. C 47 (1990) 187; \\
                 AMY Collaboration, Y.K.~Li et al.,
                 Phys. Rev. D 41 (1990) 2675; \\
                 DELPHI Collaboration, P.~Abreu et al.,
                 Z. Phys. C 73 (1997) 229.
\bibitem{dokshitzer} Yu.L.~Dokshitzer and B.R.~Webber, 
                     Phys. Lett. B 352 (1995) 451;
                     \\ Z.~Kunszt, P.~Nason, G.~Marchesini and B.R.~Webber,
                     {\em $Z$ Physics at LEP~1},
                     eds. G.~Altarelli, R.~Kleiss and C.~Verzegnassi,
                     CERN~89-08, vol.~1, p.~373.
%\bibitem{pdg} Particle Data Group, Phys. Rev. D 54 (1996) 1.
\end{thebibliography}
